\documentclass[]{article}

\usepackage[utf8]{inputenc} 
\usepackage[T1]{fontenc}    
\usepackage{url}            
\usepackage{booktabs}       
\usepackage{amsfonts}       
\usepackage{nicefrac}       
\usepackage{microtype}      

\usepackage{authblk}

\usepackage{enumitem}

\usepackage{tikz}
\usetikzlibrary{shapes.geometric, arrows}

\tikzstyle{normal} = [rectangle, rounded corners, minimum width=2.6cm, minimum height=1.0cm,text centered, draw=black]
\tikzstyle{big} = [rectangle, rounded corners, dashed, minimum width=3.0cm, minimum height=3.8cm,text centered, draw=black]
\tikzstyle{arrow} = [thick,->,>=stealth]

\usepackage{amsmath,amsfonts,amssymb,amsthm,latexsym,mathabx}
\usepackage{graphicx}
\usepackage[algo2e,ruled,vlined]{algorithm2e}
\usepackage{subcaption}
\usepackage[square,numbers]{natbib}

\newtheorem{definition}{Definition}

\DeclareMathOperator*{\argmax}{arg\,max}

\title{Reinforcement Mechanism Design, with Applications to Dynamic Pricing in Sponsored Search Auctions}
\author[1]{Weiran Shen}
\author[1]{Binghui Peng}
\author[1]{Hanpeng Liu}
\author[2]{Michael Zhang}
\author[3]{Ruohan Qian}
\author[3]{Yan Hong}
\author[3]{Zhi Guo}
\author[3]{Zongyao Ding}
\author[3]{Pengjun Lu}
\author[1]{Pingzhong Tang\thanks{Corresponding author. Email: \texttt{kenshinping@gmail.com}}}
\affil[1]{IIIS, Tsinghua University}
\affil[2]{Department of Decision Sciences and Managerial Economics, The Chinese University of Hong Kong}
\affil[3]{Baidu, Inc.}

\date{}

\begin{document}

\maketitle
\begin{abstract}

In this study, we apply reinforcement learning techniques and propose what we call reinforcement mechanism design to tackle the dynamic pricing problem in sponsored search auctions. In contrast to previous game-theoretical approaches that heavily rely on rationality and common knowledge among the bidders, we take a data-driven approach, and try to learn, over repeated interactions, the set of optimal reserve prices. We implement our approach within the current sponsored search framework of a major search engine: we first train a buyer behavior model, via a real bidding data set, that accurately predicts bids given information that bidders are aware of, including the game parameters disclosed by the search engine, as well as the bidders' KPI data from previous rounds. We then put forward a reinforcement/MDP (Markov Decision Process) based algorithm that optimizes reserve prices over time, in a GSP-like auction. Our simulations demonstrate that our framework outperforms static optimization strategies including the ones that are currently in use, as well as several other dynamic ones.

%
%
%
%
%
\end{abstract}
\section{Introduction}

In traditional markets, it is extremely difficult for firms to adjust prices when they receive more information from consumers. Partially it is because there exists a menu cost that is physically challenging to change. At the same time, even after receiving new information, it is not easy to find the best strategies to maximize profits.

In this paper, we explore the use of an AI-driven mechanism in which search engines can dynamically set prices and use the data generated in the process to maximize the profit.

Selling advertisements online through sponsored search auctions is a proven profit model for Internet search engine companies such as Google and Baidu. When a user submits a query in such a search engine, it displays, in the result page, a few advertisements alongside the organic results, both related to the query. In the backend, the query triggers an auction mechanism among all advertisers who are interested in the associated keywords. The advertisers submit bids to compete for advertising positions on the result page. The search engine then ranks the advertisers on the result page according to their bids and charges them only when some one clicks on the advertisement. 

The gold-standard mechanism in sponsored search is the well-known {\em generalized second price (GSP) auction}~\cite{edelman2007internet,varian2007position}.  The auctions allocate the best slots to the advertisers with the highest bids, second best slots to the ones with the second highest bids, and so on; and charge them based on the bids one slot below them (or the lowest price for them to maintain the current slot). Major search engines all adopt some variants of the GSP auction.

A problem with the vanilla GSP auction is that it is not revenue optimal, according the seminal theory attributed to Myerson~\cite{myerson1981optimal,riley1981optimal}. It is known that, under standard game theory assumptions, a revenue-optimal auction does not necessarily allocate the slots by the rank of their bids; it is also known that in an optimal auction, there exists a vector of advertiser-specific reserve prices that filter low bids. Over the years, a large body of literature at the interface of the economics and computation has focused on optimizing revenue of GSP auctions by incorporating insights (ranking and reserve price) from Myerson's theory~\cite{lahaie2007revenue,roberts2013ranking,ostrovsky2011reserve,milgrom2010simplified,thompson2013revenue}.

\subsection{Related works}

One objective of many works in this literature is to improve the revenue of GSP auctions~\cite{lahaie2007revenue,hartline2009simple,thompson2013revenue,shen2017practical}. When designing and analyzing these auctions, most of these works make the standard game-theoretical assumption that advertisers have a single parameter called {\em type} that indicates their maximum willingness-to-pay for a single click-through. When evaluating these auctions, these works also assume that advertisers are rational and will play according to some equilibrium. 

While these works shed lights on how to design sponsored search auctions in theory, the assumptions they make do not generally hold in the practice of keyword auctions. Most advertisers have complex private information, such as budget constraints~\cite{xu2013predicting,zhou2008budget,abrams2006revenue}, multidimensional valuation, and negative externalities~\cite{deng2011money,jehiel1996not}. Furthermore, private information such as budget may change dynamically over time and advertisers may not be able to observe all configuration parameters of the auction.

There are a few exceptions in the literature that take the initiative to design and evaluate sponsored auctions by getting rid of these assumptions. \citet{ostrovsky2011reserve} conduct large field experiments on manually setting different levels of reserve prices in sponsored search auctions and evaluate these designs. They show, with A/B tests, that by incorporating discounted Myerson's reserve prices, the search engine (Yahoo! in this case) can improve its revenue. However, it remains unclear about the long-term performance of these auctions since all these auctions are assumed to be static. It is also unclear how the ad hoc selection of the reserve prices can be improved. \citet{nekipelov2015econometrics} investigate the problem of estimating the valuations of the advertisers from their bids in the GSP auction. They get rid of the standard assumption that players must bid according to equilibrium and make a milder assumption that bidders play according to some no-regret learning strategy. They characterize the set of possible valuations given a set of bids.

In the AI community, a recent, interesting line of works aims to tackle the revenue optimization problem from a dynamic learning perspective. \citet{mohri2016learning,DBLP:conf/uai/MohriM15} apply learning algorithms to exploit past auctions as well as user features. Their algorithms mainly focus on the estimation of the underlying bid distribution, thus depending on the implicit assumption that buyers do not change their behaviors over time. \citet{mohri2015revenue,mohri2014optimal} aim to maximize advertiser revenue with strategic buyers who aim to maximize their cumulative discounted surplus. They give online pricing algorithms with desirable regret bounds. These works assume that there exists an underlying bid or value distribution for the buyers and it does not change over time.

\citet{battaglini2005long} study the Markovian consumer model in a long-term contracting setting. Their results show that even when the types at different times are highly persistent, the optimal contract is far from a static one. \citet{he2013game} and \citet{tian2014agent} also assume that buyers have the Markov property. \citet{tian2014agent} focuses on buyer behavior prediction and uses a truncated Gaussian distribution as the transition probability. Their goal is to find the best static mechanism. They also restrict their buyer model to be a linear combination of several simple behavior patterns.

\subsection{The setting: Baidu's sponsored search auction design}
In this paper, we attempt to relax these unrealistic assumptions and consider an environment in which bidders can have arbitrarily complex private information and arbitrary rationality levels that can change dynamically over time. Our goal is to design dynamic mechanisms that yield competitive revenue in practice in the long run. While the framework and algorithms proposed in this paper are applicable to search engines in general, we focus on the sponsored search auction design of Baidu, the largest search engine in China. We use Baidu as a running example throughout the paper, calibrating our model with its data.

Baidu sells 3 ad slots for most keywords and like other major search engines, Baidu runs a GSP-like auction mechanism with reserve prices to sell the slots. The bidding data yielded by different levels of reserve prices in history provides a perfect setting for us to learn how bidders react to different choices of reserve prices and the number of impressions, and the induced click-through-rates (CTRs).

\section{Preliminaries}

We consider an auction design problem in the sponsored search setting. When a user types a keyword query in a search engine, the search engine (called the seller hereafter) displays, in the result page, a few advertisements related to the keyword. We consider auctions of a single keyword, where there are $N$ bidders competing for $K$ slots. The seller allocates the slots by an auction, and each bidder $i$ reports a bid $b_i$ to the seller. A bid profile is denoted by $b=(b_1,b_2,\dots,b_N)$. We slightly abuse notations and use $b_i$ to refer to both bidder $i$ and his advertisement.

In a standard game-theoretical model, there is a single-dimensional type for each bidder that denotes the maximum amount of money that the bidder is willing to pay. However, we do not explicitly emphasize such a value in our model. The reason is two-fold: first, our model does not assume that the bidders are fully rational or rational according to some metric. Second, there are many factors that may affect bidders' bidding behavior, so explicitly define one such parameter that we cannot observe does not help much in end-to-end training. These are also the reasons why our bidder behavior model is defined over the bidders' observations and past bidding data, instead of their private information. In fact, this kind of data-driven model is not uncommon in the literature (cf. e.g.,~\cite{he2013game,xu2013predicting,pin2011stochastic}).

\subsection{Generalized second price auction}

Upon receiving a search query, the seller needs to determine a slot allocation and payment vector. Formally, a \emph{mechanism} consists of two functions $\mathcal{M}=(x, p)$, where the \emph{allocation rule} $x$ is a function $x:\mathbb{R}^N \to [0, 1]^N$, which takes as input the bid profile and outputs an $N$-dimensional vector indicating the quantity of items allocated to each bidder; and the payment rule $p$ is a function $p:\mathbb{R}^N \to \mathbb{R}^N$ that maps the bid profile to an $N$-dimensional non-negative vector specifying the payment of each bidder.

We consider the GSP auction that are widely adopted by major search engines. Suppose there are $N$ bidders competing for $K$ advertising slots. The $K$ slots have different effects of attracting user clicks (described by their CTRs). Denote by $q_k$ the CTR of the $k$-th slot and assume that $q_k$ is non-increasing with respect to the position of the slot, i.e. $q_1\ge q_2\ge \dots \ge q_K\ge 0$. Upon receiving a keyword query, the seller first collects the bid profile $b$ from the bidders. Usually, each bidder is associated with a reserve price $r_i$, which is the minimum quantity that bidder $i$ needs to bid in order to enter the auction. Denote by $b_{(i)}$ the $i$-th highest bid among those above the reserve prices. The seller then sequentially allocates the $i$-th slot to bidder $b_{(i)}$, until either the slots or the bidders run out. When bidder $b_{(i)}$'s advertisement is clicked by a user, the seller charges the bidder according to the following rule:
\begin{gather*}
p_{(i)}=
\begin{cases}
\max\left\{q_{i+1}b_{(i+1)}/q_i,r_{(i)}\right\} & \mathrm{if\ }b_{(i+1)}\mathrm{\ exists;} \\
r_{(i)} & \mathrm{otherwise.}
\end{cases}
\end{gather*}

The reserve price profile $r$ can significantly affect the revenue of the advertising platform. In this paper, we view the reserve price profile $r$ as the main parameters of the mechanism. The seller's goal is to set reserve price profiles dynamically to maximize its revenue.

\section{Our approach}
\subsection{Markov bidder model}

We assume that the bidders' behavior has time-homogeneous Markov property. Denote by $s^{(t)}_i$ and $h^{(t)}_i$ the bid distribution of bidder $i$ and the KPIs (key performance indicators) received by bidder $i$ at time step $t$. Then the bid distribution of bidder $i$ at the next time step is a function of $s^{(t)}_i$ and $h^{(t)}_i$:
\begin{gather*}
        s^{(t+1)}_i=g_i(s^{(t)}_i,h^{(t)}_i)
\end{gather*}
Such a Markov model is not uncommon in the literature, see~\cite{he2013game,battaglini2005long}. Our experiences with Baidu also indicate that the Markov model aligns with the bidders' behaviors.

\subsection{Reinforcement mechanism design}
The bids of the $N$ bidders are drawn from their bid distributions. We make the assumption that the individual bids are independent of each other. While such an assumption loses generality, it is in fact quite commonly used in the literature~\cite{DBLP:conf/uai/MohriM15,he2013game}. The joint bid distribution is
\begin{gather*}
        s^{(t+1)}=\prod\limits_{i=1}^{N}s^{(t+1)}_i=\prod\limits_{i=1}^{N}g_i(s^{(t)}_i,h^{(t)}_i) = g(s^{(t)},h^{(t)})
\end{gather*}

For simplicity, we assume that the number of daily queries of each keyword is a constant. Thus, the KPI $h^{(t)}$ is determined by both the bid distribution $s^{(t)}$ and the reserve price profile $r^{(t)}$. We view $s^{(t)}$ as the state of the seller and $r^{(t)}$ as its action, and formulate the long-term revenue maximization problem as an MDP.
\begin{definition}
        The long-term revenue maximization problem is a Markov decision process $(\mathcal{N}, S, R, G, REV(s,r),\gamma)$, where
        \begin{itemize}
        	\setlength
                \item $\mathcal{N}$ is the set of bidders with $|\mathcal{N}|=N$.
                \item $S=S_1\times\dots\times S_N$ is the state space, where $S_i$ is the set of all possible bid distributions of bidder $i$;
                \item $R=R_1\times\dots\times R_N$ is the action space, where $R_i$ is the set of all possible reserve prices that the mechanism designer can set for bidder $i$;
                \item $G=(g_1,g_2,\dots,g_N)$ is the set of state transition functions;
                \item $REV(s,r)$ is the immediate reward function that gives the expected revenue for setting reserve price profile $r$ when the state is $s$;
                \item $\gamma$ is the discount factor with $0<\gamma<1$.
        \end{itemize}
\end{definition}
The objective is to select a sequence of reserve price profiles $\{r_t\}$ that maximizes the sum of discounted revenues:
\begin{align*}
        OBJ=\sum_{t=1}^{\infty}\gamma^tREV(s_t,r_t)
\end{align*}
\subsection{Summary of the Framework}
\begin{figure}[h!]
        \centering\scalebox{.8}{
        \begin{tikzpicture}[node distance=4cm]

        \node (bidder_1) [normal] {Bidders};
        \node (eng_1) [normal, below of=bidder_1, yshift=1.8cm] {Search Engine};
        \node (algo_1) [normal, align=center, below of=eng_1, yshift=2cm] {Optimization\\ Algorithm};
        \node[text width=2.6cm, align=center] at (0,-5) {(time step $t$)};

        \node (mechanism) [big, below of=bidder_1, label=below:Mechanism, yshift=0.6cm]{};

        \node (bidder_2) [normal, right of=bidder_1] {Bidders};
        \node (eng_2) [normal, right of=eng_1] {Search Engine};
        \node (algo_2) [normal, align=center, right of=algo_1] {Optimization\\ Algorithm};
        \node[text width=2.6cm, align=center] at (4,-5) {(time step $t+1$)};

        \node (mechanism) [big, below of=bidder_2, label=below:Mechanism, yshift=0.6cm]{};

        \draw [arrow,transform canvas={xshift=-0.3cm}] (bidder_1) -- node[anchor=east,yshift=0.0cm] {bids}  (eng_1) ;
        \draw [arrow,transform canvas={xshift=0.3cm}] (eng_1) -- node[anchor=west, yshift=0.0cm] {KPIs}  (bidder_1);

        \draw [arrow] (algo_1) -- node[anchor=east, align=center] {reserve\\ prices} (eng_1);

        \draw [arrow,transform canvas={xshift=-0.3cm}] (bidder_2) -- node[anchor=east,yshift=0.0cm] {bids}  (eng_2) ;
        \draw [arrow,transform canvas={xshift=0.3cm}] (eng_2) -- node[anchor=west,yshift=0.0cm] {KPIs}  (bidder_2) ;

        \draw [arrow] (algo_2) -- node[anchor=east, align=center] {reserve\\ prices} (eng_2);

        \draw [arrow] (bidder_1) -- (bidder_2);

        \draw [arrow] (0.3,-0.8) -| (2,0);

     \draw [rounded corners, dashed] (-1.5, 0.7) -- (-1.5, -1.3) -- (2.4, -1.3) -- (2.4, -0.8) -- (5.5, -0.8) -- (5.5, 0.7) -- node[anchor=south, align=center] {Bidder model} cycle;

        \end{tikzpicture}}
        \caption{Model framework}
        \label{fig:mdp}
\end{figure}
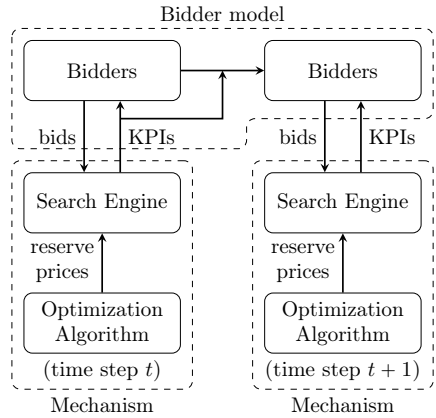

Figure~\ref{fig:mdp} shows the main framework of our model. Our model contains two parts: 1. Mechanism, where the bidders interact with the seller's action (reserve prices) and get KPIs as feedbacks; 2. Markov bidder model, which determines how bidders adjust their bids according to the KPI feedbacks.

\section{Implementation}

In this section, we describe how we implement our model in the Baidu sponsored search context and solve the optimization problem.

\subsection{Bidder model: LSTM-RNN}

In our model, each bidder is a function $g_i$ that takes as input the bid distribution $s^{(t)}_i$ and the KPIs of the bidder $h^{(t)}_i$, and outputs the bidder's bid distribution of the next time step. We discretize the distribution to 100 non-overlapping intervals. To fit the function $g_i$, we implement a standard Long Short-Term Memory (LSTM) recurrent neural network with 128 units via TensorFlow. We set the unit time step to be 1 day. The inputs of the network include KPIs of four consecutive days, the bid distributions for the bidder and also some time-specific features.

We choose the main KPIs in the system such as the number of impressions, the number of clicks and the amount of payments and take logarithm of some of the features and encode them using tile-coding. The reason of taking logarithm is based on our observation that most bidders care about the relative changes of the KPIs rather than their absolute values.



\subsection{Optimization algorithm: Monte Carlo Tree Search}

Although an optimal reserve pricing scheme exists according to the MDP theory, its exact computation is formidably costly due to the following reasons:
\begin{itemize}
        \item The possible reserve profiles of our optimization problem grows exponentially with respect to the number of the bidders;
        \item The number of future states to explore is exponential with respect to the searching depth.
\end{itemize}

We circumvent the first difficulty by restricting attentions to the keywords that contain only a few major bidders. We focus on the keywords with thin markets (few major bidders) mainly because the effect of reserve prices diminishes in thick markets anyway. To tackle the second one, we only explore possible actions for a bidder to be in a small neighborhood of the current reserve price. This restriction is also necessary for practical stability concerns, since sudden changes in reserve prices would result in sudden changes in bidders' KPIs, which would hurt the stability of the advertising platform. With these restrictions, the size of the action space is greatly reduced to a small subset. To further speed up the search, we implement the {\em Monte Carlo Tree Search (MCTS) algorithm}~\cite{khandelwal2016analysis, browne2012survey}.

The MCTS algorithm is an exploration algorithm to evaluate available action values at current state by running simulations. The MCTS algorithm maintains a tree structure, with its root representing the current MDP state. It updates the state values by repeatedly simulating available actions. Though MCTS can be replaced by any other suitable optimization algorithm in our framework, our main reason to use MCTS is that
\begin{itemize}
        \item The state space (bid distribution) has uncountably many states, therefore it would be impossible to apply the traditional MDP algorithm (value iteration or policy iteration~\cite{bellman1957dynamic}). Even discretization does not help because it still has formidable high dimensions.
        \item Though deep reinforcement learning grows fast in recent year and succeeds in numerous scenarios, it is inappropriate to train a deep neural network in our setting, i.e. deep Q-learning network(DQN)~\cite{mnih2015human} or asynchronous advantage actor-critic (A3C)~\cite{mnih2016asynchronous}. The reason is that a deep neural network depends highly on the bidder behavior model. However, the agents may leave and enter freely in a highly dynamic environment such as ours. In addition, we do not have reasonable estimations of the Q-value for each state and action, which will slow down the training process.
\end{itemize}
\subsection{Description of Algorithm}
Our MCTS algorithm starts with the root node, which represents the current state. It simulates available actions repeatedly before selecting the best one. In each simulation, the MCTS algorithm selects an action (child node) according to some selection rule and estimates the immediate reward. This procedure goes on until the some maximum depth is reached. Then the algorithm back-propagates the immediate rewards to the root node and update the corresponding long-term reward.

To estimate the expected immediate reward for an action, we simulate the corresponding auctions repeatedly and normalize the revenue according to the average number of queries for the keyword.

In general, our MCTS algorithm contains three separate parts.

\subsubsection{Selection}
We use Upper Confidence bounds for Trees (UCT) as selection strategy~\cite{kocsis2006bandit}. In UCT, we uniformly select an action until each action at a given state is select once. Then the following action is select as
$$a = \argmax_{a}\left(node.Q_{s, a} + c_p\sqrt{\frac{\text{ln}(node.n_s)}{node.n_a}} \right)$$
where $node.n_s$ is the number of times we visit $node$, $node.n_a$ is the number we select action $a$ previously, $node.Q_{s, a}$ is the current estimation of expected long term value for taking action $a$, $c_p$ can be regard as the parameter to balance exploration and exploitation. We set $c_p = \sqrt{2}$ and scale the reward to the interval $[0,1]$.
\subsubsection{Expansion and Simualtion}
During the process of exploration, more nodes are added to the tree in the expansion stage.

Upon selecting a specific action to explore, simulation is performed to derive the reward of the action and the state of the following node.
\subsubsection{Backpropagation}

Various backpropagate strategy have been develop in reinforcement learning setting~\cite{sutton1998reinforcement}. In our backpropagation algorithm, we apply SARSA$(\lambda)$~\cite{rummery1995problem}, which use $\lambda$-return to update state action value. To be more specific, the return sample is computed as
$$R_{s_t, r_t} = \sum_{n = 0}^{L - 1}w_{n}R^{n}_{s_t, a_t}$$
where $w_{n}$ satisfies 
$$w_{n} = \left\{\begin{matrix}
(1 - \lambda)\lambda^{n} & 0 \leq n < L - 1\\
\lambda^{L - 1} &L - 1\\
\end{matrix}
\right.$$




Our back-propagation algorithm are stated in Algorithm~\ref{algo:bp}.
\begin{algorithm2e}[!h]
        \caption{Back-propagation Algorithm for $\lambda$-return}
        \KwIn{Sample Path: $path$ }
        $q = 0$ \\
        \For{$t = L - 1$; $t > 0$; $t \leftarrow t - 1$ }
        {
                $(node, a, r) \leftarrow path[t]$\\
                $node.n_s \leftarrow node.n_s + 1,$\\
                $node.n_a \leftarrow node.n_a + 1$\\
                $q \leftarrow q + r,  \delta_{Q} \leftarrow q - node.Q_{s,a}$\\
                $node.Q_{s,a} \leftarrow node.Q_{s,a} + (\delta_{Q}/node.s_a)$\\
                $q \leftarrow (1 - \lambda)\max_{a^{\star}| node.n_{a^{\star} \neq 0}} [node.Q_{s, a^{\star}}] +  \lambda q$
        }
\label{algo:bp}
\end{algorithm2e}

\section{Simulations}
\label{sec:exp}
In this section, we describe how we compare different algorithms by simulation based on real data from Baidu.
We extract 8 months' bidding data from Baidu, and selected 400 keywords\footnote{Our dataset is considerably larger than in most papers in the literature. For example, ~\cite{nekipelov2015econometrics} conduct experiments based on 1 week's data from 9 bidders and the dataset for simulations in~\cite{lahaie2007revenue} contains only 1 keyword.} that meet the following conditions:

\begin{itemize}
        \item The number of daily queries for the keyword is large and stable (with small variance).
        \item The most part (at least $80\%$) of the revenue of the keyword is contributed by at most 3 bidders.
\end{itemize}

\begin{figure}[h!]
        \centering
        \begin{subfigure}[t]{0.22\textwidth}
                \centering
                \includegraphics[width=\textwidth]{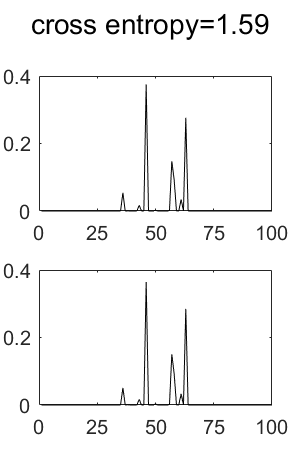}
        \end{subfigure}
        \begin{subfigure}[t]{0.22\textwidth}
                \centering
                \includegraphics[width=\textwidth]{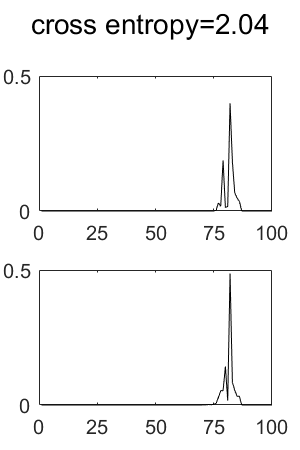}
        \end{subfigure}
        \begin{subfigure}[t]{0.22\textwidth}
                \centering
                \includegraphics[width=\textwidth]{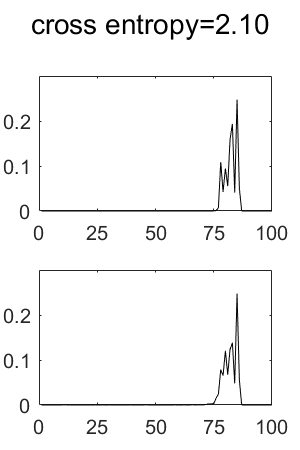}
        \end{subfigure}
        \begin{subfigure}[t]{0.22\textwidth}
                \centering
                \includegraphics[width=\textwidth]{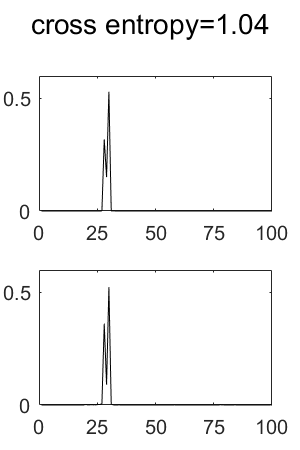}
        \end{subfigure}
        \caption{Prediction results for 4 selected bidders. Each sub-figure contains two distributions, with the upper one being the actual distribution and the lower one being the prediction. The cross entropy of each sub-figure is shown on top.}
        \label{figure: lstm result}
\end{figure}

For each keyword, we only focus on the 3 major bidders and ignore others. For each bidder, we trained a recurrent neural network using the 8 months' data and use cross entropy as the performance indicator for the network. We use 10\% of all data as the test set. The average cross entropy among all bidders and all test instances is $1.67$. Some selected test instances are listed in Figure~\ref{figure: lstm result}.

We explore possible reserve prices for the bidder to be $95\%$, $100\%$ and $105\%$ times the current reserve price for the bidder. We set $\lambda = 0.8$ and the search depth to be 5 in our optimization algorithm. In the selection step, we restrict the number of exploration to be 5000. In the expansion step, to estimate the revenue at the selected node, we simulate the auction 5 million times and compute the average revenue as the per-impression revenue of each keyword.

We set the initial reserve price to be $p = \argmax_{b} b(1-F(b))$ where $F(b)$ is the current bid distribution. We call this reserve price static optimal, since this price maximizes the revenue if the bidders do not change their bids. Several algorithms are compared by our simulations:
\begin{itemize}
        \item STATIC\_OPT: Always use the initial reserve.
        \item GREEDY: In each round, we randomly choose a bidder and change his reserve price by $-5\%$ and simulate auctions for the next period, if the revenue for the next period goes up, then use this reserve price. Otherwise, change his reserve price by $+5\%$. Notice this method can been seen as a simplified version of coordinate gradient descend(ascend) method.
        \item BAIDU: 
        Current reserve prices used by Baidu.
        \item STATIC\_50: 50 cents as the reserve prices for all bidders, regardless of bid distribution.
\end{itemize}

We also compare the effect of different frequencies of changing reserve prices by setting the time step $\Delta t$ in the expansion step of the optimization algorithm\footnote{This time step is not necessarily equal to the time step for training the Markov bidder model. We can always simulate bidder behaviors day by day but change the reserve every several days}. Clearly, changing the reserve prices too frequently can affect the stability of the platform and thus is not desirable. In this simulation, we only compare the performance of our framework.

\subsection{Results and analysis}

\begin{figure}[h!]
	\centering
	\begin{subfigure}[t]{0.4\textwidth}
		\centering
		\includegraphics[width=\textwidth]{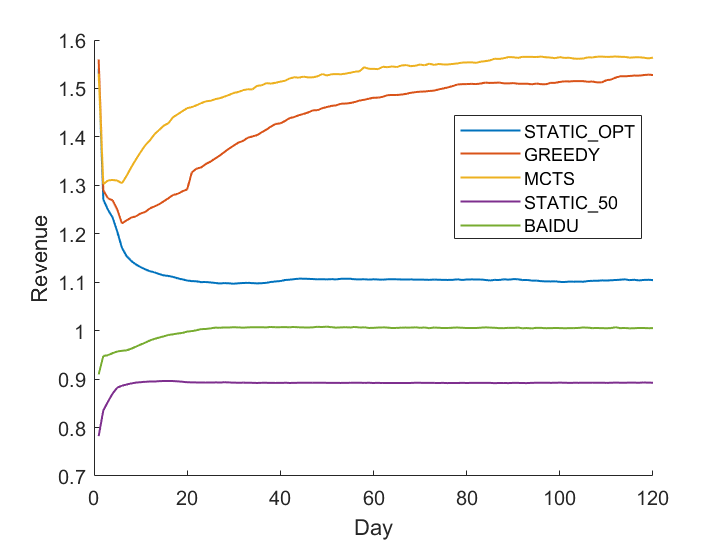}
		\caption{Performance of different strategies}
		\label{fig:exp1}
	\end{subfigure}
	\begin{subfigure}[t]{0.4\textwidth}
		\centering
		\includegraphics[width=\textwidth]{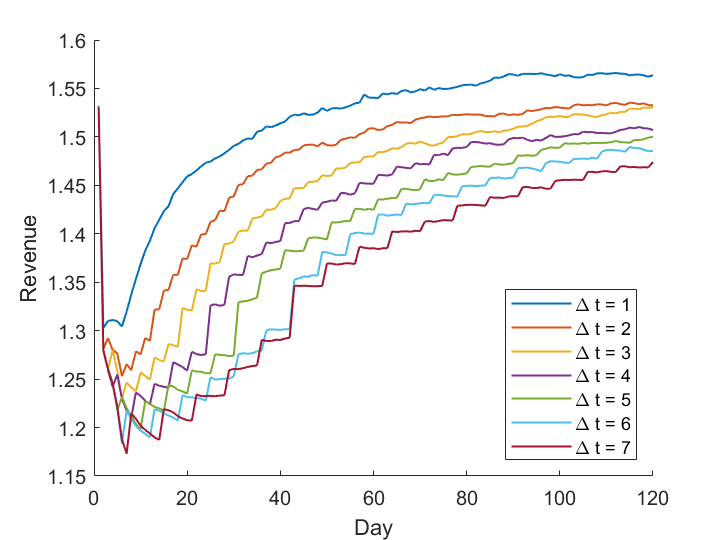}
		\caption{Effect of the frequency of changing reserve prices}
		\label{fig:exp2}
	\end{subfigure}
\end{figure}

In the first simulation, we change the reserve every day (time step $\Delta t=1$) in our MCTS algorithm, and compare it with other strategies mentioned above. We simulate 120 days for each strategies. The results of the simulations are shown in Figure~\ref{fig:exp1}. Revenue is normalized with the converged value of BAIDU. The figure shows that
\begin{itemize}
        \item Our dynamic strategy outperforms all other static strategies (STATIC\_OPT, BAIDU, STATIC\_50) as well as the dynamic strategy GREEDY;
        \item The BAIDU curve converges rapidly within just few days. The reason that the curve still has a convergence phase is that our simulation is a simplified version of Baidu's auction mechanism. 
        \item The STATIC\_OPT curve undergoes a rapid rise on the first day and then followed by a steep fall, also converges after two weeks. 
\end{itemize}

Besides, the simulation also reveals some interesting facts about bidder behaviors:
\begin{itemize}
        \item All aggressive pricing schemes gain high revenue immediately and drops significantly later. This phenomenon is intrinsic for our dataset, since all the bidders undergo mild pricing mechanism previously due to moderate choice of reserve prices. The sudden change in reserve price could make huge immediate reward, but once bidders are aware of the change and respond accordingly, less revenue can be extracted.
        \item Although STATIC\_OPT could beat mild mechanism like BAIDU and STATIC\_50, its long term revenue is not as promising as the short term. However, by adopting dynamic mechanism, we can gradually increase daily revenue.
        \item The simulation shows that with more involved optimization algorithm (such as MCTS) and accurate bidder model, we could achieve the best performance and gain higher revenue in the long run.
\end{itemize}

In the second simulation, we compare the effect of the frequency of changing reserve prices. The results are shown in Figure~\ref{fig:exp2}. We also simulated 120 days for each $\Delta t$. the figure indicates that the larger $\Delta t$ is, the more revenue it can extract, and the more quickly it converges. The revenue of $\Delta t=7$ is about several percent small than that of $\Delta t=1$, Comparing Figure~\ref{fig:exp1} and ~\ref{fig:exp2}, we can see that the performance GREEDY algorithm is almost the same as the MCTS algorithm with $\Delta t=3$.
%
%

\section{Conclusion}
In this paper, we propose a dynamic pricing framework, which we call reinforcement mechanism design, that combines reinforcement learning and mechanism design. Our framework does not depend on unrealistic assumptions adopted by most theoretical analyses. Interestingly, our framework uses a data-driven approach to solve a theoretical market-design problem.

Our framework contains two main parts: the bidder-behavior model and the optimization algorithm. The optimization algorithm finds the optimal mechanism parameters for each step repeatedly. In each round, the algorithm estimates the future objectives by simulating the auctions with the bidder-behavior model.

We apply our framework to the sponsored search setting and assume Markov bidder behavior. The model uses an RNN for the bidder model and an MCTS algorithm to solve for the optimal reserve prices. Our simulations with real bidding data from a major search engine in China show that our framework can dramatically improve the revenue compared to other static and dynamic strategies.

\bibliographystyle{plainnat}
\bibliography{ref}
\end{document}